\documentclass{svproc}
\usepackage{url}

\usepackage{multicol}
\usepackage{amsmath}
\usepackage{graphicx}
\usepackage[table,xcdraw]{xcolor}
\usepackage{subcaption}
\usepackage{mwe}

\usepackage[round, sectionbib]{natbib}   % omit 'round' option if you prefer square brackets
\begin{document}
\mainmatter 

\title{Exploring shareability networks of probabilistic ride-pooling problems}

\titlerunning{Topology in ride-pooling} 

\author{Michał Bujak\inst{1} \and Rafał Kucharski\inst{1}}
\authorrunning{Michal Bujak and Rafal Kucharski}
\institute{Jagiellonian University, Faculty of mathematics and computer science, Prof. Stanisława Łojasiewicza 6, 30-348 Kraków, Poland, \\
contact email: \email{michal.bujak@doctoral.uj.edu.pl}}
\maketitle              % typeset the title of the contribution

\begin{abstract}
Travellers sharing rides in ride-pooling systems form various kinds of networks. While the notions of the so-called shareability graphs, has been in the core of many ride-pooling algorithms, so far they have not been explicitly analysed. Here, we introduce and examine four kinds of networks resulting from ride-pooling problems. 

We use 147 NYC taxi requests from 2016 pooled into attractive shared-rides with our utility based ExMAS algorithm and explore resulting shareability networks. To cover unknown individual properties of pooling travellers, we run 1000 replications of probabilistic pooling process, resulting in richer representation of weighted graphs instrumental to reveal complex network structures. 

Our findings reveal substantial differences between network structures and topologies. Properties of ride-pooling networks may be further analysed to better understand and improve ride-pooling systems.

\end{abstract}

\section{Introduction}
Ride-pooling, where travellers share a vehicle to reach their destinations, is becoming an increasingly popular urban mobility alternative. 
Thanks to pooling, co-travellers can reduce their travel costs, mobility platforms (like Uber and Lyft) can more efficiently utilise their fleets and cities can reduce congestion.

Naturally, travellers sharing their rides form a graph structures, already identified in the seminal works of \cite{santi2014quantifying} and \cite{alonso2017demand}. 
While the so-called shareability networks are central to solve the ride-pooling problems they have not been hitherto explicitly analysed.

Ride-pooling, from the networks perspective, can be seen as linking two or more nodes. Node is a traveller, or more precisely her/his single trip request (tuple of origin, destination and request time). The link is formed at the two levels. First, when all the co-travellers agree to travel together (in our case this means that the ride is attractive for all the co-travellers). Second, when they actually travel together (when each traveller is uniquely assigned to a given ride). This two interpretations of links, yields two different network structures (second one being a subgraph of the first one). We will refer to the former representation as \emph{shareability}, i.e. potential to travel together, and the latter as \emph{matching}, i.e. actually travelling together. Moreover, the links can be formed twofold: first, when travellers (nodes) are linked directly, and second when we introduce a new kind of node: the ride, which can be seen as a vehicle trip during which co-travellers travel together. Linking co-travellers with a ride results in a bipartite network. We will denote the former as \emph{simple networks} and latter as \emph{bipartite networks}. This two levels of twofold distinctions yield four kinds of networks that we explore in this paper.

Moreover, to represent the probabilistic nature of underlying pooling decisions made by travellers, we propose the probabilistic version of pooling algorithm. By adding the random noise to the utility formulas of ride-pooling algorithm we introduce the behavioural non-determinism to the pooling problem. When replicated multiple times we obtain rich and meaningful structures of weighted graphs which well estimate probabilistic graphs. 
With those we can observe both the stable matches with strong links, as well as components where pooling compositions can vary substantially between replications. %from day to day.

Such weighted shareability graphs, consistent with the actual travel behaviour, can be useful not only for a deeper understand of the still challenging ride-pooling problems, but also to identify communities in ride-pooling networks, improve attractiveness of pooling services or control the virus spreading in ride-pooling networks (like in \cite{kucharski2021virus}).

\subsection{Background}
The potential of ride-pooling to improve urban mobility has triggered a bursting stream of research (\cite{agatz2011dynamic}; \cite{chan2012ridesharing}; \cite{furuhata2013ridesharing}; \cite{Tachet2017}; \cite{bischoff2017city}), which led to an algorithmic breakthrough in the seminal works of \cite{santi2014quantifying} and \cite{alonso2017demand}. First, in \cite{santi2014quantifying}, the so-called shareability network was introduced along with a methodology to match travellers in pairs. Further exploited in \cite{alonso2017demand} with the complete algorithm to efficiently match incoming requests with available vehicles by sequentially adding new co-travellers into vehicles with empty seats and adjusting their routes. By applying cutoffs on maximal detours and delays the pooled rides are acceptable for travellers and by minimising the service costs (vehicle hours) in the matching the solution becomes optimal for the supply side (platform and/or drivers). In parallel, behavioural research has made advancements in better understanding users’ so-called willingness to share (\cite{kang2021pooled}; \cite{lavieri2019modeling}; \cite{alonso2021determinants}; \cite{lazarus2021pool}). Such, utility-driven approaches, not only put the traveller in the centre but also allow to incorporate inherent non-determinism of travellers' decision processes.

Hitherto shareability and matching structures have been only introduced. We formalise and analyse different methods of presenting their topologies. Furthermore, we introduce random noise, hence the novel non-deterministic approach to ride-pooling. Based on replications, we introduce weighted graphs which estimate the probabilistic graphs. We analyse their evolution and stability.

\begin{section}{Methodology}

\begin{subsection}{Ride-pooling algorithm}
ExMAS \citep{kucharski2020exmas} is an offline algorithm which matches the travellers into a shared ride if it is more attractive than travelling alone. More precisely, travellers are assigned to the shared rides if individual utility of the shared ride exceeds a non-shared ride utility for each of the co-travellers. Value of the utility depends on the individual- and system-dependent variables with the following  formula, given as the difference between shared ride and non-shared ride:
\begin{equation}\label{equation1}
    U = U^s - U^{ns} = \beta^c \lambda l + \beta^t(t - \beta^s(\hat{t} + \beta^d\hat{t}^p)) +\epsilon > 0,
\end{equation}
where $U$, $U^s$, $U^{ns}$ denote respectively utility gain due to sharing, utility of shared ride and utility of non-shared ride. $\lambda$ stands for discount for sharing a ride and is controlled by the system operator. $\beta^c$, $\beta^t$, $\beta^s$ $\beta^d$ are the behavioural parameters: cost sensitivity, value of time, sharing discomfort and delay sensitivity, respectively. $t$ and $\hat{t}$ stand for travel time of non-shared and shared ride, respectively, $\hat{t}^p$ is a delay associated with pooling and $\epsilon$ is a random term. 

We say that a group of travellers can share a ride, if for each of them the individually perceived utility of the ride is positive. Those rides contribute to the set of attractive (feasible) rides.
ExMAS identifies such rides hierarchically, first exploring pairwise shareable trips, which are then extended to triples, quadruples, etc. until no cliques can be found in shareability networks.
On the set of identified attractive shared rides we solve the assignment problem in which travellers are uniquely assigned to a rides, formulated as a matching on a bipartite graph aimed to minimise the total distance travelled. From now on, we will refer to the set of feasible, attractive rides identified with ExMAS as \emph{shareability} and the optimal solution as a \emph{matching}.
Full details of ExMAS algorithm are available in \cite{kucharski2020exmas}.

Hitherto, the deterministic version of ExMAS has been used, with $\epsilon = 0$.Here, we decided to adopt a probabilistic model in which $U$ follows a normal distribution (probit model). Formally, probability that traveller $i$ wants to participate in ride $j$ is described as follows, $\mathbf{P}(t_i \in r_j) = \Phi_{0, \sigma} (U_s - U_{ns})$, where $\Phi_{0, \sigma}$ stands for CDF of the normal distribution with mean $0$ and standard deviation $\sigma$. For the sake of simplicity, we omit in $U$ additional indexes, however keeping in mind that $U$, $U_{ns}$ and $U_s$ are individually calculated for each traveller for each ride. $\Phi_{0, \sigma}$ is a strictly increasing bijection, hence, we can write:
\begin{align*}
    \mathbf{P}(t_i \in r_j) = \Phi_{0, \sigma}(U_s - U_{ns}) > v \\
    U_s - U_{ns} > \Phi_{0, \sigma}^{-1}(v).
\end{align*}
Aforementioned inequalities yield a simple way to draw values to estimate the distributions of non-deterministic ride-pooling. %The random term corresponds to individual bias towards ride-pooling. Hence, we assume that within each simulation, the random term is traveller-specific, drawn only once. 
Each replication yields simulated values of the random term, hence giving explicit values of utility per each set of travellers.

\end{subsection}

\begin{subsection}{Construction of networks}\label{construction_networks}

Both finding the shareability and the matching can be represented with networks. The most straightforward representation is a network, where travellers constitute nodes, linked when they share a ride (row one in tab. 1). In such \emph{shareability} network (first column in tab.1), travellers are linked if they find it attractive, whereas in the \emph{matching} network (second column) if the ride is part of the optimal solution. While the latter is the actual realisation of ride-pooling, i.e. linked travellers actually travel together, the former reflects the possibility to travel together (which may not be realised due to other constrains).

Such network representation is, however, incomplete. From the shareability graph, we cannot conclude if, for example, a triangle means that the three travellers can share a ride together or whether they are just pairwise shareable (Figure \ref{fig:pairs_shareability_single}). In order to avoid this information loss, an alternative representation if often proposed. Both a shareability and matching network can be formulated as bipartite graphs, where nodes are not only travellers but also the rides (vehicles). Traveller is connected to a ride in the bipartite shareability network if for her/him the ride is attractive (Figure \ref{fig:bipartite_shareability_single}), in case of bipartite matching network travellers are linked only to rides being part of the optimal solution (Figure \ref{fig:bipartite_matching_single}). This leads to the four distinct network structures, as summarised in the Table \ref{tab:types_network} and illustrated with Figure \ref{fig:single_nets}.

\begin{table}
\begin{center}
\begin{tabular}{c|c|c}
     \textbf{network} &  \textbf{nodes} & \textbf{links}  \\ \hline
     shareability & travellers & attractive pooling (between travellers)\\
     matching &   & optimal matching  (between travellers)\\ \hline
    bipartite shareability & travellers  & attractive pooling (between travellers and rides)  \\
    bipartite matching &   and rides & optimal matching (between travellers and rides) \\ \hline
\end{tabular}
% \begin{tabular}{c|c|c|c}
%      &  shareability & matching & nodes & links \\ \hline
%      travellers & A & B & travellers  \\
%      bipartite &  C & D & travellers and rides (vehicles)
% \end{tabular}
\end{center}
\caption{\label{tab:types_network} \footnotesize{Four networks of ride pooling.}}
\end{table}

\begin{figure*}
    \centering
    \begin{subfigure}[b]{0.52\textwidth}
        \centering
        \includegraphics[width=\textwidth]{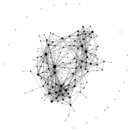}
        \caption[Network2]%
        {{\small Shareability network}}    
        \label{fig:pairs_shareability_single}
    \end{subfigure}
    \hfill
    \begin{subfigure}[b]{0.45\textwidth}  
        \centering 
        \includegraphics[width=\textwidth]{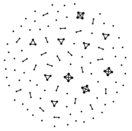}
        \caption[]%
        {{\small Matching network}}    
        \label{fig:pairs_matching_single}
    \end{subfigure}
    \vskip\baselineskip
    \begin{subfigure}[b]{0.46\textwidth}   
        \centering 
        \includegraphics[width=\textwidth]{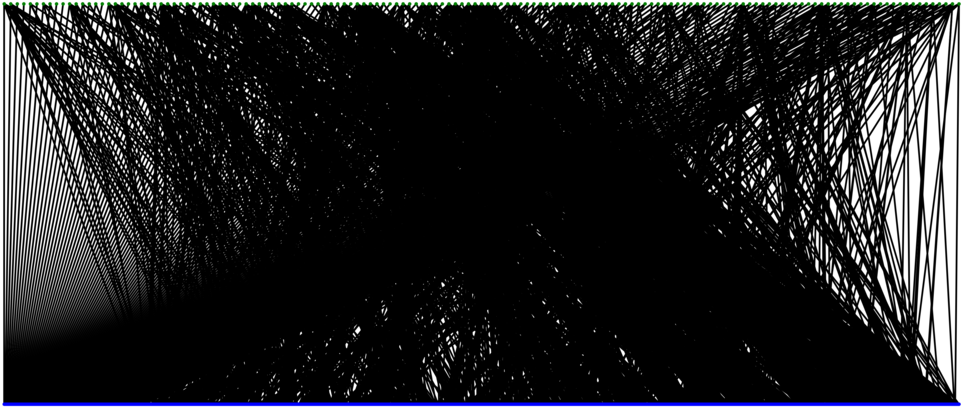}
        \caption[]%
        {{\small Bipartite shareability}}    
        \label{fig:bipartite_shareability_single}
    \end{subfigure}
    \hfill
    \begin{subfigure}[b]{0.46\textwidth}   
        \centering 
        \includegraphics[width=\textwidth]{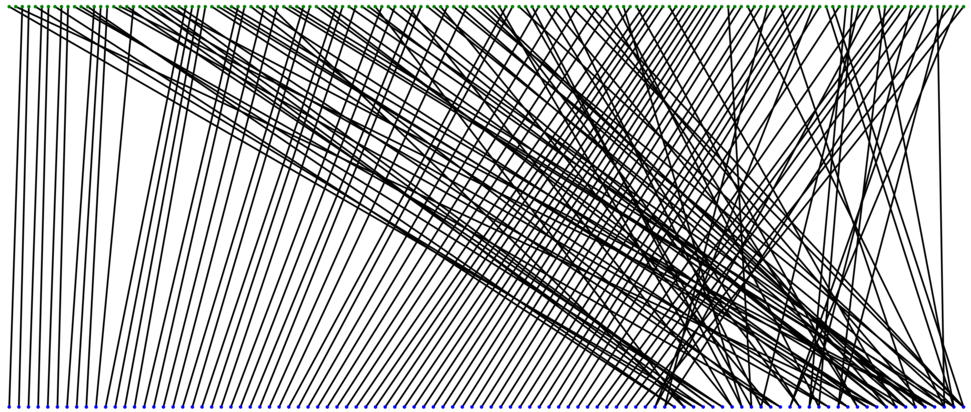}
        \caption[]%
        {{\small Bipartite matching}}    
        \label{fig:bipartite_matching_single}
    \end{subfigure}
    \caption[ Figures single rep ]
    {\small Four representations of networks in ride-pooling on a single replication from the case of 147 pooling travellers in NYC. The shareability network (a) is highly connected, many nodes (travellers) are linked with multiple other potential co-travellers. However, when travellers are uniquely assigned to vehicles in the matching network (b) the structure becomes dramatically more sparse. The bipartite forms of those two networks have similar structures. Travellers (top) are linked with rides (bottom). Among multiple potentially feasible links (c) only few remain the part of actual matching (d).} 
    \label{fig:single_nets}
\end{figure*}

\end{subsection}

\end{section}
\begin{section}{Experiment}

We experiment with trips actually requested within half an hour in New York City on January 2016. Each of 147 requests links origin with destination at a given time (as illustrated on Figure \ref{fig:demand}). We assumed the fare $\beta^c$ is $1.5$ \$/km, sharing discount $\lambda$ is $30\%$, value of time $\beta^t=9$\$/h and willingness to share $\beta_s=1.3$ (most of which based either on empirical observations or actual system values). We set pooling parameters to be constant among all the replications (deterministic part) and assume the random term $\epsilon$ of utility $U$ to follow standard normal distribution.
\begin{figure}
    \centering
    \includegraphics[width=.85\textwidth]{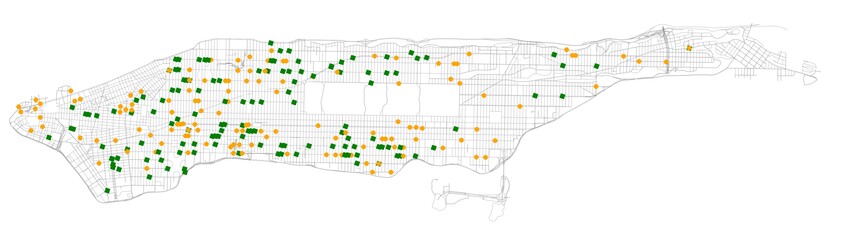}
    \caption{Demand dataset for experiments: 147 trip requests from Jan 2016 in Manhattan. Green dots are origins, orange destinations.}
    \label{fig:demand}
\end{figure}

It is important to note that even though we can explicitly calculate utility of a ride for any group of travellers hence probability of corresponding links, there is no easy way to determine probability in the matching. Thus, to observe different realisations of the random term $\epsilon$, we have performed $1000$ replications. For each, we have constructed networks such as shown in the Figure \ref{fig:single_nets}.

\begin{subsection}{Results}
 First, we look at the properties of ride-pooling networks and their stability across the replications. Aggregated results are presented in the Table \ref{tab:aggregated_struct}. Some of the networks’ properties remain constant by their definition. For example, the number of nodes in the simple networks is equal to the number of travellers. Similarly, the number of links in bipartite matching. However, investigating different structural characteristics reveals interesting patterns. Relatively big standard deviation in the number of nodes and edges in the bipartite shareability suggests that there is indeed a significant structural variance introduced by the random noise. Low deviation in the bipartite shareability network of the average degree of a ride suggests that introduced noise has a subsidiary impact on the average occupancy of vehicles among feasible rides. For the same type of network, average degree of traveller nodes has a significant deviation. The fact implies that there is a substantial group of rides which balance on the verge of attractiveness. Hence, minor disturbance introduced by the noise can have a significant impact on the number of feasible rides for a traveller. Furthermore, relating to the variance of average degree of nodes in the simple shareability network, we may conclude that noise is pronounced both in potential sharing between two travellers and for the higher degree rides. Relative size of the greatest component shows significant variance for the bipartite shareability network. Considering the fact that the maximal value is $1$, while the average is $0.96$, we can conclude that often the greatest component becomes much smaller. That may suggest forming at least two big components which are not always connected. Further investigation could prove communities presence within the components.

\begin{table}[]
\resizebox{\textwidth}{!}{%
\begin{tabular}{c|c|c|c|c|c|c}
\rowcolor[HTML]{FFFFFF} 
\textbf{}                       & \textbf{nodes} & \textbf{edges} & \textbf{degree} & \textbf{deg.rides} & \textbf{deg.travellers} & \textbf{greatest   component} \\ \hline
\rowcolor[HTML]{F5F5F5} 
shareability & 147 (0) & 470 (47) & 6,394 (0,638) &       &               & 0,853 (0,027) \\
\rowcolor[HTML]{FFFFFF} 
matching       & 147 (0) & 61 (6)   & 0,831 (0,082) &       &               & 0,030 (0,004) \\
\rowcolor[HTML]{F5F5F5} 
bipartite shareability & 1164 (166)     & 2306 (497)     & 3,922 (0,287)   & 2,249 (0,121)           & 15,692 (3,384)     & 0,961 (0,11)                  \\
\rowcolor[HTML]{FFFFFF} 
bipartite   matching & 250 (3) & 147 (0)  & 1,175 (0,15)  & 1 (0) & 1,425 (0,043) & 0,022 (0,002) \\ \hline

\end{tabular}
}

\caption{\label{tab:aggregated_struct}\footnotesize{General properties and their variability in $1000$ replications (std in brackets) for four introduced graph structures.} }
\end{table}

We can now introduce a weighted graphs which aggregate realisations of the probabilistic graph. For the simple networks, nodes remain constant. Hence, we link two nodes if for any of the replications they were connected. The weight of the link is exactly the number of occurrences of the link in the experiment. We can introduce bipartite networks in nearly the same way. The only difference is that for rides nodes, we must take union of all rides which occurred in any replication. The introduced graph structures are presented in the Figure \ref{fig:prob_nets}. Width of the link proportionally depends on the weight.

\begin{figure*}
    \centering
    \begin{subfigure}[b]{0.48\textwidth}
        \centering
        \includegraphics[width=\textwidth]{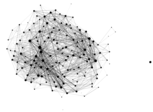}
        \caption[Network2]%
        {{\small Shareability network}}    
        \label{fig:mean and std of net14}
    \end{subfigure}
    \hfill
    \begin{subfigure}[b]{0.48\textwidth}  
        \centering 
        \includegraphics[width=\textwidth]{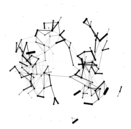}
        \caption[]%
        {{\small Matching network}}    
        \label{fig:mean and std of net24}
    \end{subfigure}
    \vskip\baselineskip
    \begin{subfigure}[b]{0.46\textwidth}   
        \centering 
        \includegraphics[width=\textwidth]{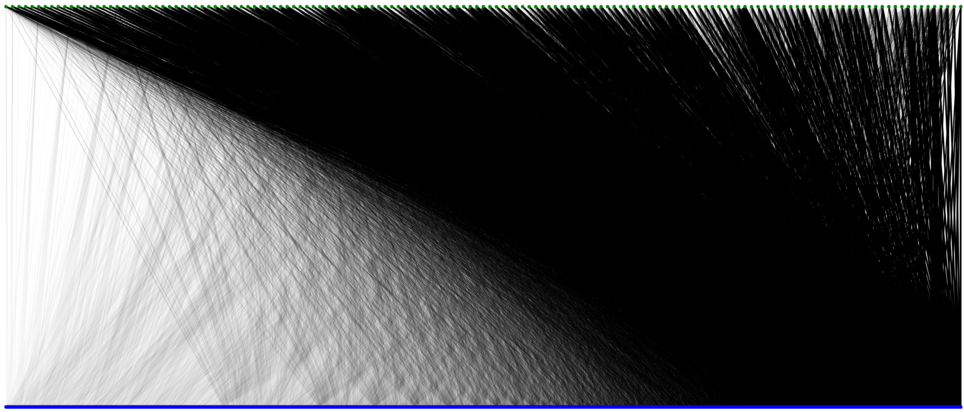}
        \caption[]%
        {{\small Bipartite shareability}}    
        \label{fig:mean and std of net34}
    \end{subfigure}
    \hfill
    \begin{subfigure}[b]{0.46\textwidth}   
        \centering 
        \includegraphics[width=\textwidth]{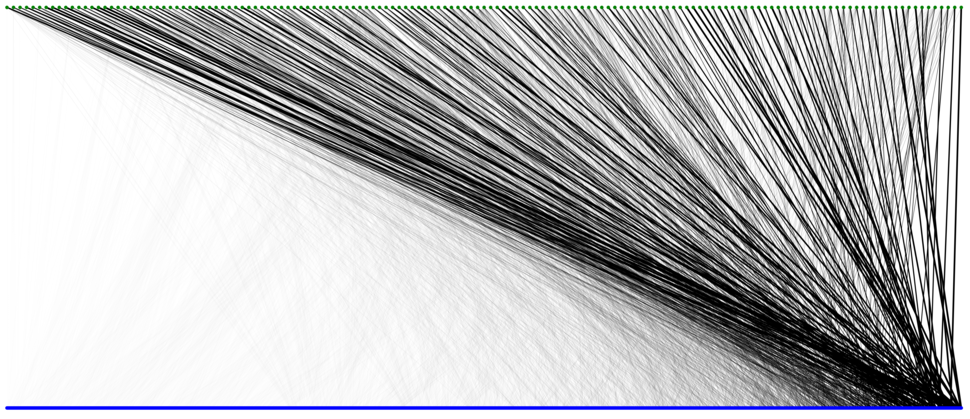}
        \caption[]%
        {{\small Bipartite matching}}    
        \label{fig:mean and std of net44}
    \end{subfigure}
    \caption[ Figures single rep ]
    {\small Weighted shareability networks. After several replications of the stochastic matching process the network wiring evolves. The potential shareability network (a) is becoming even more connected, similarly to the matching, where some travellers are stably matched with high probability, while others build a broad set of potential co-travellers (b). Similar trends can be seen in the bipartite versions of the networks (c and d).} 
    \label{fig:prob_nets}
\end{figure*}

Intrigued by the structure of graphs in the Figure \ref{fig:prob_nets}, we investigated the evolution of bipartite shareability networks. Figure \ref{fig:evo_nets} presents weighted graphs constructed after a certain part of the simulation. We have found that even in the latter replications, new ride nodes appeared. Those nodes usually correspond to high order rides, where travellers are matched thanks to positive realisations of the random noise. Considering the fact that the solution is optimal with respect to saved vehicle hours, attractive high order rides are likely to be assigned in the final matching. Nonetheless, we can see that there are certain rides which remain attractive and assigned in nearly every replication.

\begin{figure*}
    \centering
    \begin{subfigure}[b]{0.196\textwidth}
        \centering
        \includegraphics[width=\textwidth]{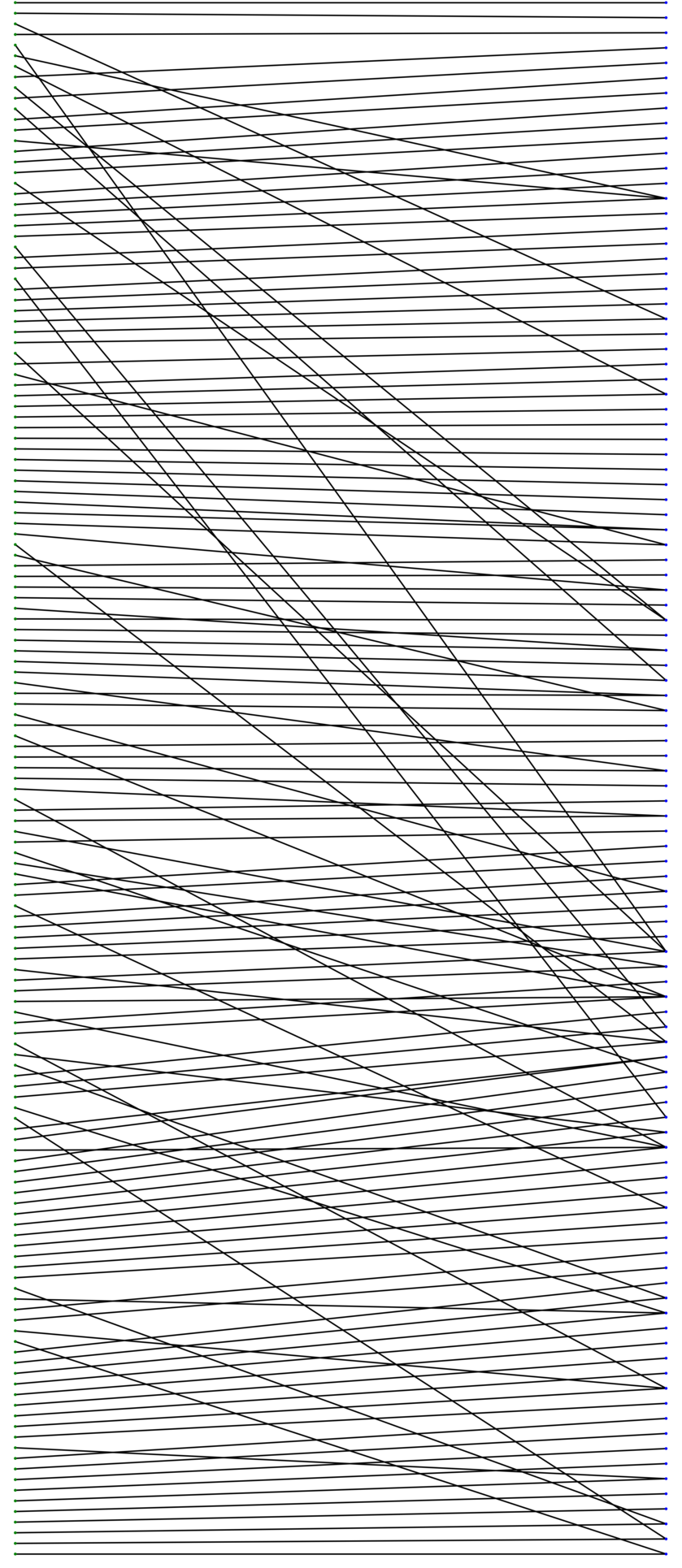}
        \caption[]%
        {{\small 1st step}}    
        \label{fig:mean and std of net14}
    \end{subfigure}
    \hfill
    \begin{subfigure}[b]{0.189\textwidth}
        \centering
        \includegraphics[width=\textwidth]{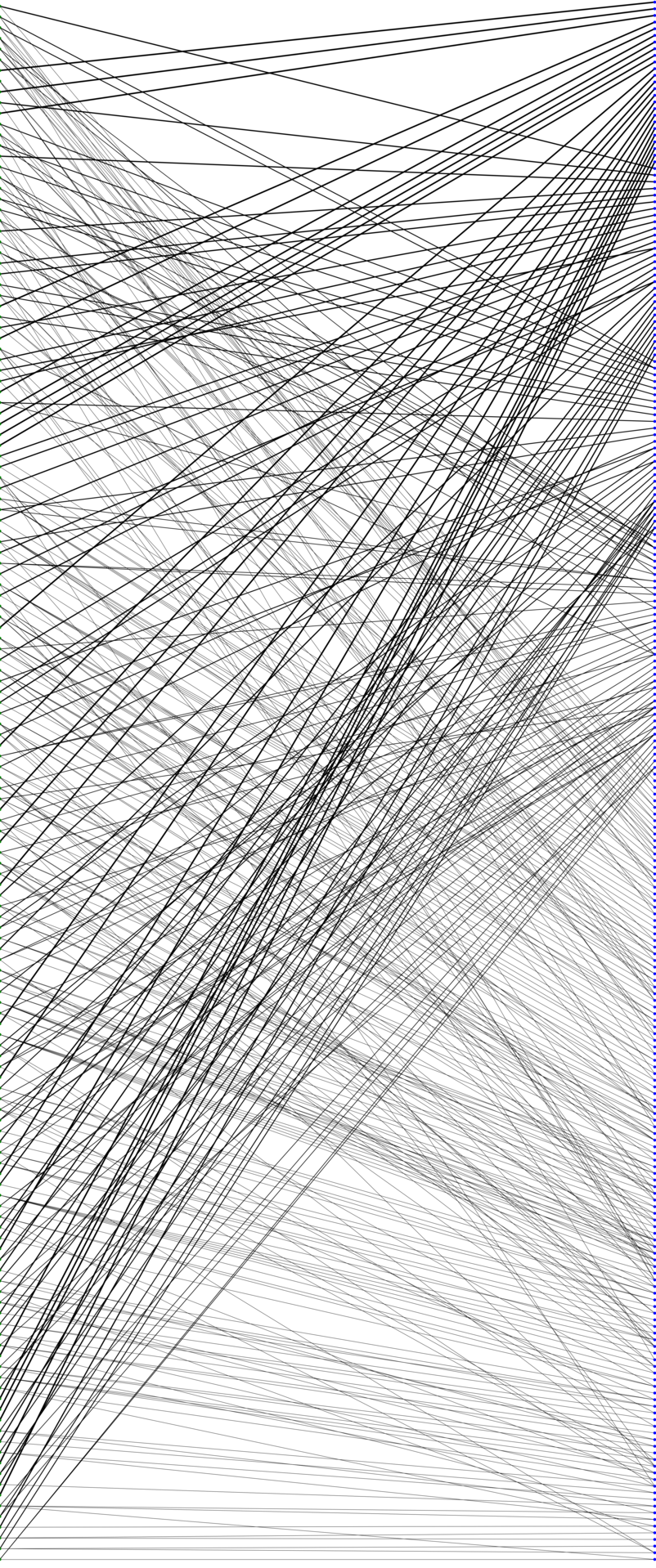}
        \caption[]%
        {{\small 5 rep.}}    
        \label{fig:mean and std of net14}
    \end{subfigure}
    \hfill
    \begin{subfigure}[b]{0.189\textwidth}  
        \centering 
        \includegraphics[width=\textwidth]{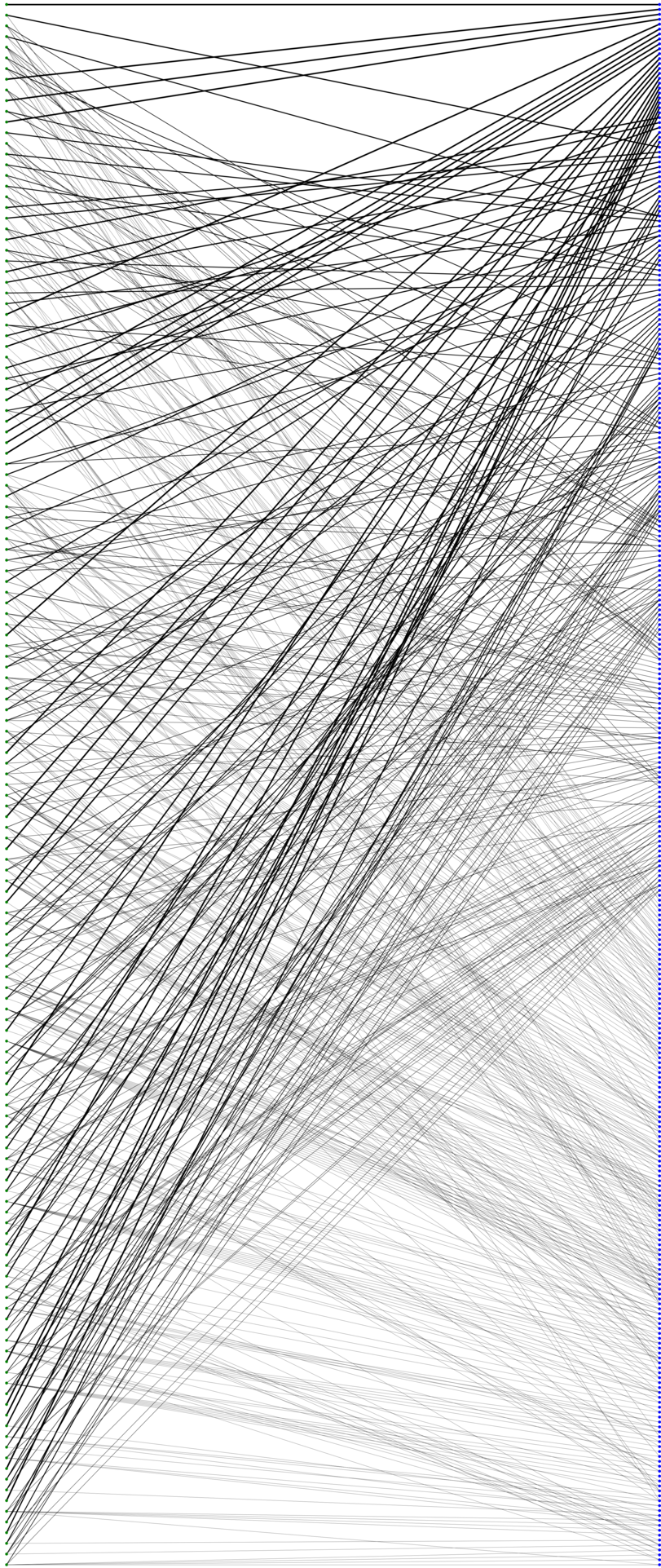}
        \caption[]%
        {{\small 10 rep.}}    
        \label{fig:mean and std of net24}
    \end{subfigure}
    % \vskip\baselineskip
    \hfill
    \begin{subfigure}[b]{0.19\textwidth}   
        \centering 
        \includegraphics[width=\textwidth]{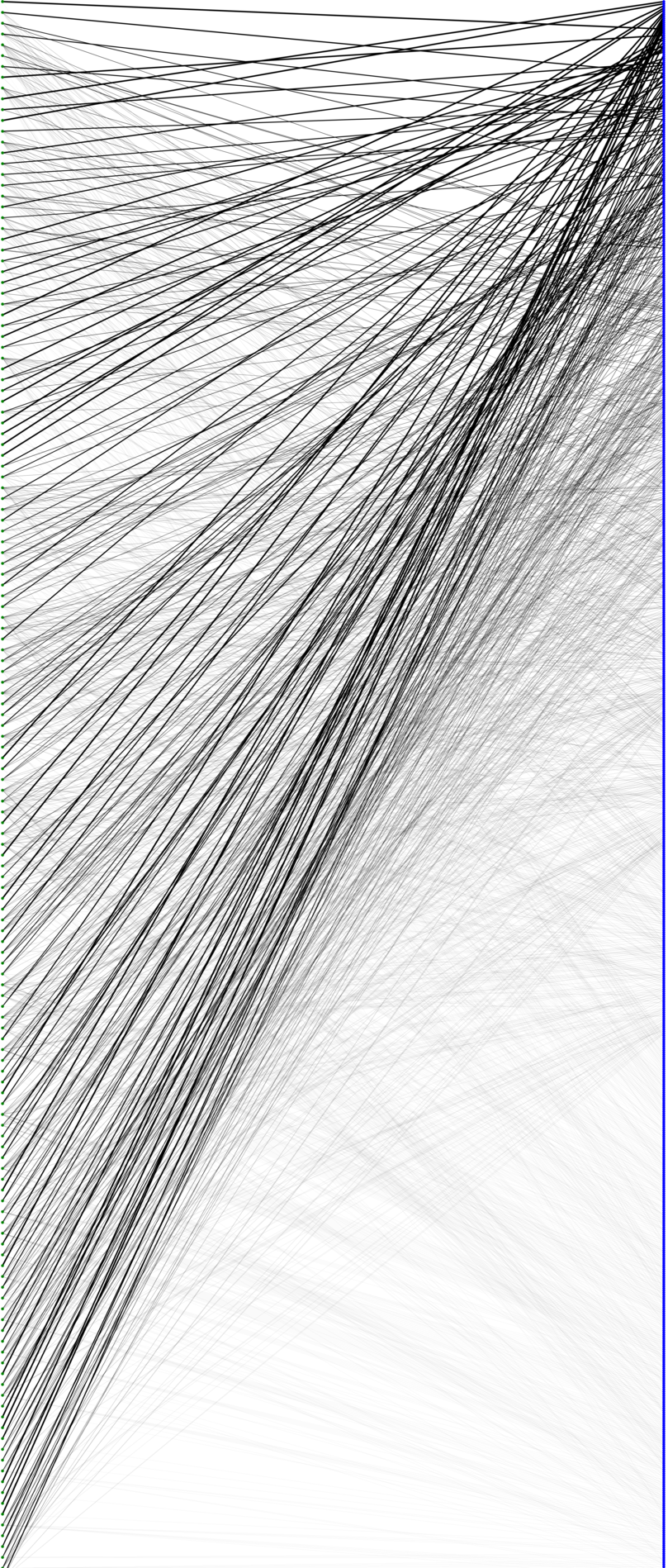}
        \caption[]%
        {{\small 100 rep.}}    
        \label{fig:mean and std of net34}
    \end{subfigure}
    \hfill
    \begin{subfigure}[b]{0.19\textwidth}   
        \centering 
        \includegraphics[width=\textwidth]{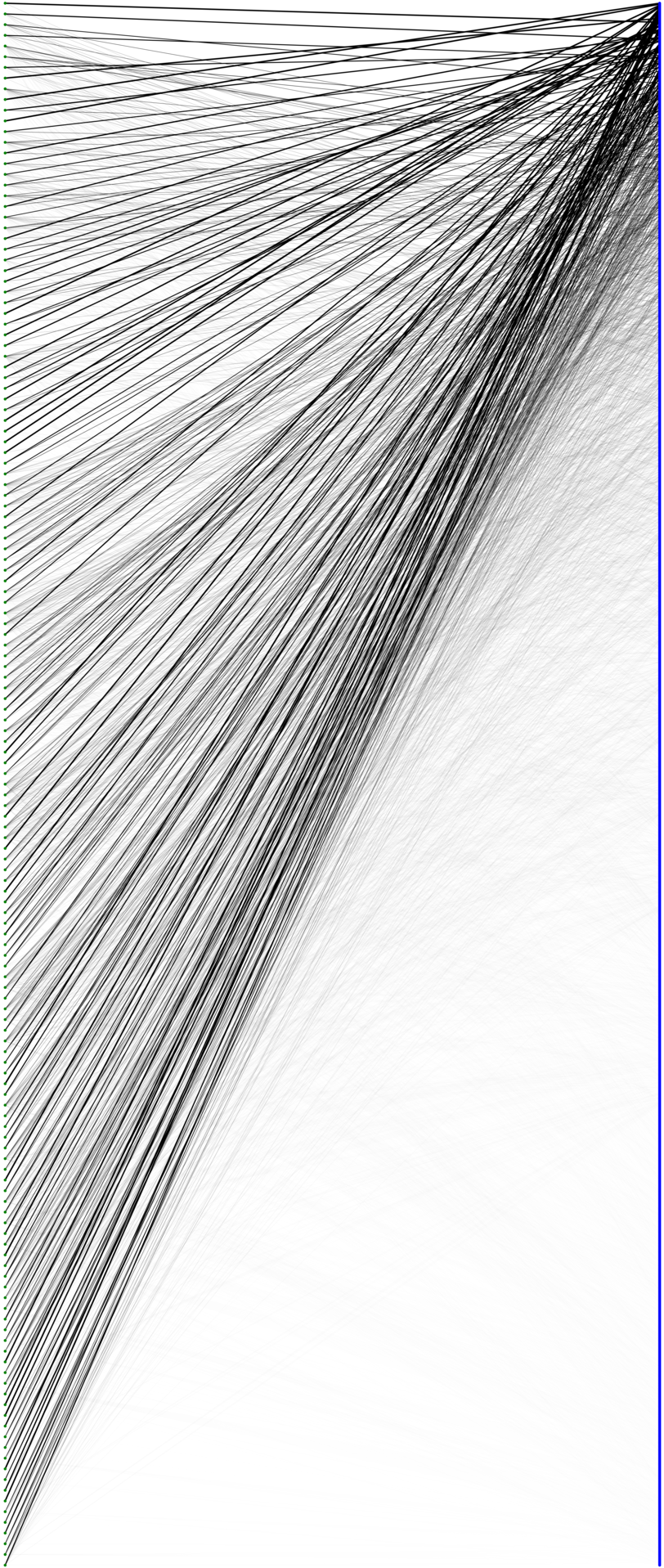}
        \caption[]%
        {{\small 900 rep.}}    
        \label{fig:mean and std of net44}
    \end{subfigure}
    \caption[ Figures single rep ]
    {\footnotesize{Evolution of the bipartite matching between travellers (left) and rides (right) (panel d on figs. \ref{fig:single_nets} and \ref{fig:prob_nets}). The number of rides (on the right) as well as number of connections increases. Some highly attractive relations become stronger, while others appear rarely, with low probability.}}
    \label{fig:evo_nets}
\end{figure*}

% \begin{figure*}
%     \centering
% \includegraphics[width=.85\textwidth]{plots/evolution_matching_bipartite.jpg} 
% \caption{\small Evolution of the matching between travellers (left) and rides (right) in the bipartite form. The number of rides (on the right) as well as number of connections increases. Some highly attractive relations become stronger, while others appear rarely, with low probability.} 
%     \label{fig:mean and std of nets}
% \end{figure*}

To gain more understanding of evolution in the Figure \ref{fig:evo_nets}, we investigated properties of the (weakly) growing weighted matching graph in the non-bipartite form. Figure \ref{fig:frame_props} resents how the average node degree and relative size of the greatest component grow. Those properties have a direct, physical interpretation and application. Average degree (blue) of the node reflects the average number of pairwise distinct co-travellers that a traveller meets within the $1000$ realisations. Size of the greatest component (red) gives an upper boundary for spreading in the network starting in one point (to obtain exact results we should translate our partial results to temporal graph).

\begin{figure*}[h]
    \centering
\includegraphics[width=.85\textwidth]{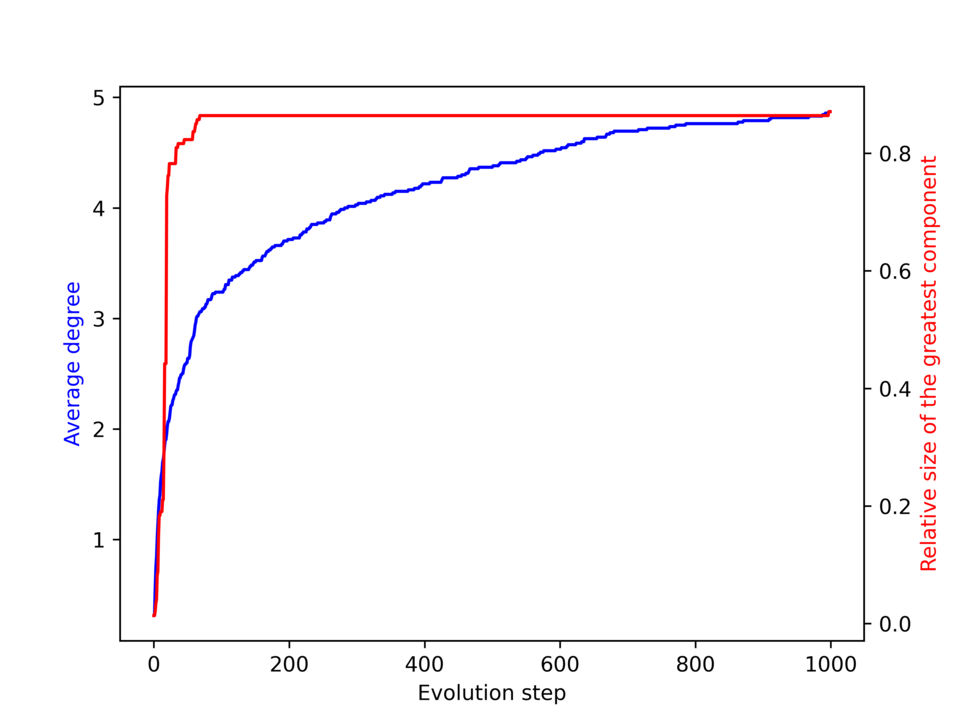}
\caption{\label{fig:frame_props} \small Evolution of graph properties along the replications. As we explore the probabilistic nature of ride-pooling, the graph properties evolve. The greatest component of the matching graph (network B in the Figure \ref{fig:prob_nets}) quickly reaches stable value and only slightly evolves after 900 days of evolution. Average degree also increases, yet slower, gradually increasing until the end of the simulation. } 
\end{figure*}
\end{subsection}

\end{section}
\section{Conclusions}
We have presented four ways to represent the networks resulting from ride-pooling problems. In the first type of the networks, each node corresponds to a traveller. Second representation incorporates also rides as nodes, enabling an exhaustive representation for the feasibility network.

 Considering individual preferences not captured by a deterministic model, we introduce a random noise assigned to each traveller. As a result, within each replication we got different realisations of the random variables. We analyse the introduced variability in the context of structural properties of networks. Weighted graphs representing the estimated probability of occurrence of a link yield new form of networks in ride-pooling. Tracking evolution of the graphs gave us a better understating of the stability of the underlying structures. In future works, we plan to focus on the communities and analysing spreading patterns in the network.

\paragraph{Acknowledgements} This research was funded by National Science Centre in Poland program OPUS 19 (Grant Number 2020/37/B/HS4/01847).

\bibliography{ref}

\end{document}